\documentstyle[12pt]{article}
\def\lap{\hbox{~{\lower -2.5pt\hbox{$<$}}\hskip -8pt\raise 
-3.5pt\hbox{$\sim$}}}
\def\gap{\hbox{~{\lower -2.5pt\hbox{$>$}}\hskip -8pt\raise 
-3.5pt\hbox{$\sim$}}}
\def\apg{\hbox{{\raise -2.5pt\hbox{$>$}}\hskip -8pt\lower -2.5pt\hbox{$\sim$}}}
\def\apl{\hbox{{\raise -2.5pt\hbox{$<$}}\hskip -8pt\lower -2.5pt\hbox{$\sim$}}}
\begin{document}

\centerline{\bf REDUCTION OF THE WAVEPACKET: HOW LONG DOES IT TAKE?$^*$}\footnote{*This paper 
was published in 1984 as a Los Alamos report LAUR 84-2750, and was 
widely circulated at the time. It is based on the talks given by the author 
in the Spring of 1984 at the ITP, UCSB workshop on {\it Quantum Noise}, where
the key result relating decoherence time in quantum Brownian motion, Eq. (1),
was first announced, and at the NATO ASI {\it Frontiers of Nonequilibrium
Quantum Statistical Physics}, held in Santa Fe, June 3-16, 1984. It was 
eventually published in the proceedings of that meeting, NATO ASI vol. B135,
pp. 145-149, G. T. Moore and M. O. Scully, eds. (Plenum, 1986). It is 
reproduced here in the original version, with only a few typos corrected.} \\

\vspace{.05 in}
\begin{center}
W.H. Zurek \\
Theory Division, MS B213\\
Los Alamos National Laboratory \\
Los Alamos, New Mexico 87545 \\
and \\
Institute for Theoretical Physics \\
University of California \\
Santa Barbara, California 93106
\end{center}
\vspace{.05 in}
\begin{abstract}
We show that the ``reduction of the wavepacket'' caused by the interaction 
with the environment occurs on a timescale which is typically many orders of
magnitude shorter than the relaxation timescale $\tau$. In 
particular, we show that in a system interacting with a ``canonical'' 
heat bath of
harmonic oscillators decorrelation timescale of two pieces of the 
wave-packet separated by $N$ thermal de Broglie wavelengths is
approximately $\tau/N^2$. Therefore, in the classical limit $\hbar 
\rightarrow 0$ dynamical reversibility $(\tau \rightarrow \infty)$ is
compatible with ``instantaneous'' coherence loss.
\end{abstract}

\vspace{.25 in}
\noindent
{{INTRODUCTION}}
\vspace{.15 in}

It is sometimes argued that observables of macroscopic objects which 
obey, to a good approximation,
\emph{reversible} classical dynamics -- i.e. their relaxation 
timescale t is, for all practical purposes, infinite -- could not have
lost coherence and become ``classical'' due to the interaction with 
the environment through environment-induced superselection.$^{1-5}$
For, the reasoning goes, relaxation rate is the measure of the 
strength of the coupling with the environment. In particular, when 
$\tau
\rightarrow \infty$ one can neglect dissipation of energy. 
Consequently, one should be equally justified in neglecting any 
influence of
the environment. We show that this argument is fallacious in an 
example of a free particle interacting with the environment of quantum
oscillators in the high-temperature weak coupling limit. In 
particular, we show that the coherence between two pieces of the 
wave-packet
$\Delta x$ apart is lost on a \emph{decorrelation timescale} $\theta$ 
which is typically
\begin{equation}
\theta = \tau\left [ \left(\hbar / \sqrt{4mkT} \right) / \Delta x 
\right]^2 \ \ .
\end{equation}
Here, $m$ is the mass of the particle, $k$ is Boltzmann's constant, 
and $T$ is temperature. For ``canonical'' classical
systems $(m \sim 1g, \ \ T \sim 300^o K)$ and standard 
``macroscopic'' separations $\Delta x \sim 1 cm , \  \theta/\tau \sim 
10^{-40}$.
Moreover, in the classical limit, $\hbar \rightarrow 0 , \ 
\theta/\tau \rightarrow 0$.  This enormous disparity between the two
timescales can be regarded as the explanation of the apparent 
``instantaneous'' collapse of the state vector of macroscopic objects,
including distinguishable (i.e. separated by many \emph{de Broglie 
wavelength}*\footnote{*The more popular definition of
\emph{thermal de Broglie wavelength} is $\lambda^2_T = h^2/2\pi mkT$. 
It differs by a factor $\sqrt{\pi / 2} \  (\lambda^2_{dB} = (2/
\pi) \lambda^2_T)$ from the de Broglie wavelength $\lambda_{dB}$ we 
shall use here.} $\lambda_{dB} = \hbar/\sqrt{4mkT}$) outcomes of
measurements performed by a classical apparatus on a quantum system.

\vspace{.25 in}
\noindent
{\bf{DECORRELATION OF A ``FREE PARTICLE''}}
\vspace{.15 in}

Consider an otherwise free particle of mass $m$ interacting with the 
environment of many harmonic oscillators via the Hamiltonian:
\begin{equation}
H_{INT} = x \sum c_iq_i \ \ .
\end{equation}
Above, $x$ is the coordinate of the free particle while $q_i$ are the 
coordinates of harmonic oscillators. This interaction
Hamiltonian was used extensively in many earlier discussions of 
relaxation,$^{7,8}$ and, more recently it is being used in 
calculations of
dephasing in a harmonic oscillator.$^9$

We note that $H_{INT}$, Eq.~(1), commutes with the position 
observable of the free particle:
\begin{equation}
\left[H_{INT}, x \right] = 0 \ \ .
\end{equation}
Therefore, position can be regarded as \emph{pointer 
observable},$^{1,3,5}$ measured continuously by the environment of 
harmonic
oscillators. In the absence of the self-Hamiltonian:
\begin{equation}
H_0 = - \left( \hbar^2 / 2m \right) \left( \partial / \partial x \right)^2
\end{equation}
$x$ would be a constant of motion. One would then expect combined 
system-environment state vector to evolve from an initial, 
uncorrelated
state
$$
|\phi_0 \rangle = |\psi \rangle | \epsilon \rangle
$$
into the time dependent, correlated state:
$$
|\phi_t \rangle \propto \int dx |\psi(x) \rangle | \epsilon \rangle \ \ .
$$
Tracing out an environment after it has performed idealized, perfect 
``measurement'' -- i.e. after states of the environment are
correlated with different positions become orthogonal, $\langle 
\epsilon_x|\epsilon_y \rangle \sim \delta(x - y)$ -- yields, for the
system, the density matrix:
\begin{equation}
\rho \propto \int dx  |\psi(x)|^2  | x \rangle \langle x |
\end{equation}
This density matrix is diagonal in the pointer basis $\{|x\rangle\}$.

In the more realistic case of finite $H_0$ the density matrix $\rho$ 
will not achieve perfect diagonalization, Eq. (5). Rather, it will 
have a
finite correlation length $\sim \lambda_{dB}$. Moreover, the 
distribution will become uniform, $\langle x|\rho |x\rangle$ = 
const., on a
relaxation timescale. The estimate of the timescales of these two 
processes can be obtained from the effective master equation for the
free particle. We shall use it in the form given by Caldeira and 
Leggett$^7$. Its three consecutive terms correspond to the von 
Neumann's
equation for the density matrix of a free particle, to the 
dissipation with viscosity $\eta = 2 m \gamma$, and to the 
fluctuating force
responsible for Brownian motion:
\begin{eqnarray}
\dot{\rho} & = & \left\{\left( i \hbar /2m \right) \left(\partial^2 / 
\partial x^2 - \partial^2 / \partial y^2 \right) - \gamma(x-y)
\left( \frac{\partial}{\partial x} - \frac{\partial}{\partial 
y}\right) \right. \nonumber \\
& & - \left.\ ^{}_{}\left(2 m \pi k T/ \hbar^2 \right) (x - y)^2 \right\}\rho
\end{eqnarray}
To compare relaxation and decorrelation timescales we consider an 
initial wavepacket of half-width $\delta$. As we shall argue in the 
next
section, this half-width will be typically of the order of the de 
Broglie wavelength. We now suppose that the initial wavepacket has 
been
``split,'' coherently, into two pieces, $|\alpha \rangle$ and $|\beta 
\rangle$, so that the free particle is described by the wave function:
\begin{equation}
|\psi \rangle = \left(|\alpha \rangle +  |\beta \rangle\right) / \sqrt{2} \ \ .
\end{equation}
Here we assume for simplicity:
\renewcommand\theequation{8a}
\begin{equation}
\langle x|\alpha \rangle  =  (2 \pi \delta^2)^{-1/4} \exp\left[ -(x - 
\Delta x / 2)^2 / 4 \delta^2\right] \ \ ,  \\
\end{equation}
\renewcommand\theequation{8b}
\begin{equation}
\langle x|\beta \rangle  =  (2 \pi \delta^2)^{-1/4} \exp\left[ -(x - 
\Delta x / 2)^2 / 4 \delta^2\right] \ \ .  \\
\end{equation}
The resulting initial density matrix
\renewcommand\theequation{9}
\begin{equation}
\rho = | \psi \rangle \langle \psi |
\end{equation}
has, in the position representation, four extremes. Two of them occur 
on the diagonal: (1) $x = y = \Delta x/2$; (2) $x = y = -\Delta x/2$.
They are the maxima of $|\langle x | \alpha \rangle |^2$ and 
$|\langle x | \beta \rangle |^2$. In addition, there are off-diagonal
maxima of $\langle x | \alpha \rangle\langle x | \beta \rangle$ and 
of its Hermitian conjugate which lie at: (3) $x = -y = \Delta x/2$;
(4) $x = -y = -\Delta x/2$. The size of these off-diagonal maxima 
provides a measure of the coherence between $|\alpha \rangle$ and
$|\beta \rangle$.

The rate of change of the diagonal terms due to the interaction with 
the environment can be estimated by calculating,
from Eq. (6),
\renewcommand\theequation{10a}
\begin{equation}
\tau^{-1} = \langle \alpha_t |\dot{\rho}| \alpha_t \rangle \cong - 
(\gamma / 2) \langle \alpha_t | (x - y)^2| \alpha_t \rangle
\left (1/\delta^2 + 1/\lambda^2_{dB}\right)
\end{equation}
Here $|\alpha_t \rangle = \exp (-iH_0t/\hbar)|\alpha\rangle$ was used to 
separate out the evolution due to the environment from the evolution
induced by the self-Hamiltonian $H_O$. Similarly, the rate of change 
of the off-diagonal term is:
\renewcommand\theequation{10b}
\begin{equation}
\theta^{-1} = \langle \alpha_t |\dot{\rho}| \beta_t \rangle \cong - 
(\gamma / 2) \langle \alpha_t | (x - y)^2| \beta_t \rangle
\left (1/\delta^2 + 1/\delta^2_{dB}\right)
\end{equation}
The key and only difference between the two rates is then the size of 
the factor $\langle(x-y)^2\rangle$. For the diagonal terms it is given by
\renewcommand\theequation{11a}
\begin{equation}
\langle \alpha_t | (x - y)^2| \alpha_t \rangle = \delta^2 \sim 
\lambda^2_{dB} \ \ .
\end{equation}
For the off-diagonal elements, it is, on the other hand
\renewcommand\theequation{11b}
\begin{equation}
\langle \alpha_t | (x - y)^2| \beta_t \rangle = (\Delta x)^2 \ \ .
\end{equation}
Therefore, the ratio of the two rates is
\renewcommand\theequation{12}
\begin{equation}
\tau/\theta  = (\Delta x/\delta)^2 \sim (\Delta x/\lambda_{dB})^2 \ \ .
\end{equation}
in accord with Eq. (1). For ``macroscopic'' values of $\Delta x$, 
$m$, and $T$, this ratio is enormous and enforces environment-induced
superselection. It is worth pointing out that qualitative conclusions 
of our discussion are in accord with more elaborate path integral
treatment of the harmonic oscillator, given recently by Caldeira and 
Leggett.$^9$

\vspace{.25 in}
\noindent
{\bf{DISCUSSION: A CLASSICAL LIMIT?}}
\vspace{.15 in}

In the previous section we have shown that when $\delta \sim 
\lambda_{dB}$, decorrelation is much more rapid than relaxation. The 
purpose
of this section is to justify why, in the practical context, the 
assumption $\delta \sim \lambda_{dB}$ is natural. Moreover, we shall
briefly point out consequences of the disparity between the two 
timescales for the interpretation of quantum mechanics.

Let us first consider a classic example of measurement, patterned 
after the one discussed by von Neumann.$^{10}$
We couple the measured system, initially in a state $|\phi\rangle$, with the 
free particle measuring apparatus, so that their total Hamiltonian is
\renewcommand\theequation{13}
\begin{equation}
H  = H_{SYSTEM} + H_0 - i\hbar \Delta x \delta( t - t_0)P(\partial / 
\partial x) \ \ .
\end{equation}
Here $P$ is the measured operator which we assume has 0 and 1 as the 
eigen-values, while $x$ is
the position of the free particle which will record the outcome of 
the measurement.

Just before the observation the state of the apparatus must be deter- 
mined with the accuracy better than $\Delta x$.  If the free
particle apparatus is already in contact with the heat bath of 
temperature $T$, as discussed previously, then the measurement of its
position with some accuracy $\sigma$, $\Delta x \gg \sigma \gg
\lambda_{dB}$, will be a typical, sufficient preparation.  Therefore, 
the
apparatus will be left in an incoherent mixture of $n = \sigma / 
\lambda_{dB}$ wavelets.  Such inexhaustive measurements may be not 
only
``realistic,'' but also advantageous, as the resulting mixture will 
spread slower than the pure wavepacket of comparable width.$^{11}$
In the course of the interaction at $t = t_0$, each of the de 
Broglie-sized wavepackets will be split into an ``unmoved'' $|\alpha
\rangle$ portion, and into the shifted one:  $\exp (-i \Delta x 
\partial / \partial x) | \alpha \rangle = |\beta \rangle$. 
Therefore,
immediately after the observation, the state of the combination 
(system - free particle apparatus) is, in effect, described by a 
mixture of
terms of the form:
\renewcommand\theequation{14}
\begin{equation}
|\Upsilon \rangle  = | \alpha \rangle ( 1 - P) | \phi \rangle + | \beta 
\rangle P| \phi \rangle
\end{equation}
with all the $|\alpha \rangle$ contained within $\sigma$.  Now the 
analysis of the decay of the pure state $|T \rangle $ into the
density matrix of the form
\renewcommand\theequation{15}
\begin{equation}
  \rho  = | \alpha \rangle \langle \alpha | ( 1 - P) | \phi \rangle 
\langle \phi |(1 - P) + | \beta \rangle \langle \beta | P| \phi 
\rangle
\langle \phi | P
\end{equation}
can be conducted in accord with the discussion of the previous 
section.  In particular, $\delta \sim \lambda_{dB}$ will apply as 
long as
the resolution $\sigma$ of the measurement which prepares 
free-particle apparatus is worse than $\lambda_{dB}$.  Moreover, even 
if $\sigma < \lambda_{dB}$, our qualitative conclusions still hold, as in 
that case decorrelation will be even more rapid.

The most intriguing corollary of our discussion is, perhaps, the 
possibility that in the classical limit of $\hbar \rightarrow 0$ the
relaxation timescale may approach infinity,
\renewcommand\theequation{16}
\begin{equation}
  \tau \rightarrow \infty
\end{equation}
which allows the system to follow reversible, Newtonian dynamics, and 
yet the decorrelation timescale will remain arbitrarily short,
or, indeed, it may approach zero:
\renewcommand\theequation{17}
\begin{equation}
  \theta \rightarrow 0 \ .
\end{equation}
We regard this limit as a \emph{true classical limit}: Not only does 
it allow classical Newton's equations of motion, but it also prevents
long-range quantum correlations, by imposing the environment-induced 
superselection.$^{1,2}$

It is worth stressing that the loss of coherence and the accompanying 
``irreversibility'' is a consequence of the deliberate tracing out
of the environment, which disposes of the mutual information,$^5$ and 
not of the approximations involved in the derivation of Eq. (6). This
is particularly clearly demonstrated by the analogous results 
obtained by path-integral methods for harmonic oscillators.$^9$

I would like to thank Amir Caldeira and Dan Walls for discussions on 
the subject of this paper. This research was supported by the
National Science Foundation under Grant No. PHY77-27084, supplemented 
by funds from the National Aeronautics and Space Administration.

\end{document}